\input harvmac
\overfullrule=0pt
\parindent 25pt
\tolerance=10000
\input epsf

\newcount\figno
\figno=0
\def\fig#1#2#3{
\par\begingroup\parindent=0pt\leftskip=1cm\rightskip=1cm\parindent=0pt
\baselineskip=11pt
\global\advance\figno by 1
\midinsert
\epsfxsize=#3
\centerline{\epsfbox{#2}}
\vskip 12pt
{\bf Fig.\ \the\figno: } #1\par
\endinsert\endgroup\par
}
\def\figlabel#1{\xdef#1{\the\figno}}
\def\encadremath#1{\vbox{\hrule\hbox{\vrule\kern8pt\vbox{\kern8pt
\hbox{$\displaystyle #1$}\kern8pt}
\kern8pt\vrule}\hrule}}

 \def\frac#1#2{{#1\over #2}}

 \def\s{\sqrt}

 \def\al{\alpha'}
 \def\de{\partial}

 \def\lr{\leftrightarrow}
 \def\f {\frac}
 \def\ti{\tilde}
 \def\ap{\alpha}

 \def\ddd{\cdot\cdot\cdot}
 
 \def\la{\langle}
 \def\lb{\rangle}

\lref\AKKT{ S.~Y.~Alexandrov, V.~A.~Kazakov and I.~K.~Kostov,
``Time-dependent backgrounds of 2D string theory,'' Nucl.\ Phys.\
B {\bf 640}, 119 (2002) [arXiv:hep-th/0205079].
}

\lref\tdmat{ D.~J.~Gross and N.~Miljkovic, ``A Nonperturbative
Solution Of D = 1 String Theory,'' Phys.\ Lett.\ B {\bf 238}, 217
(1990);
E.~Brezin, V.~A.~Kazakov and A.~B.~Zamolodchikov,
``Scaling Violation In A Field Theory Of Closed Strings In One
Physical Dimension,'' Nucl.\ Phys.\ B {\bf 338}, 673 (1990);
P.~Ginsparg and J.~Zinn-Justin, ``2-D Gravity + 1-D
Matter,'' Phys.\ Lett.\ B {\bf 240}, 333 (1990).
}

\lref\DK{P.~Di Francesco and D.~Kutasov, ``Correlation functions
in 2-D string theory,'' Phys.\ Lett.\ B {\bf 261}, 385 (1991);
``World sheet and space-time physics in two-dimensional
(Super)string theory,'' Nucl.\ Phys.\ B {\bf 375}, 119 (1992)
[arXiv:hep-th/9109005].
}

\lref\MV{ J.~McGreevy and H.~Verlinde, ``Strings from tachyons:
The c = 1 matrix reloaded,'' JHEP {\bf 0312}, 054 (2003)
[arXiv:hep-th/0304224];

J.~McGreevy, J.~Teschner and H.~Verlinde,
``Classical and quantum D-branes in 2D string theory,''
JHEP {\bf 0401}, 039 (2004)
[arXiv:hep-th/0305194].
}

\lref\KMS{ I.~R.~Klebanov, J.~Maldacena and N.~Seiberg, ``D-brane
decay in two-dimensional string theory,'' JHEP {\bf 0307}, 045
(2003) [arXiv:hep-th/0305159].
}

\lref\TT{T.~Takayanagi and N.~Toumbas, ``A matrix model dual of
type 0B string theory in two dimensions,'' JHEP {\bf 0307}, 064
(2003) [arXiv:hep-th/0307083].
}

\lref\six{ M.~R.~Douglas, I.~R.~Klebanov, D.~Kutasov,
J.~Maldacena, E.~Martinec and N.~Seiberg, ``A new hat for the c =
1 matrix model,'' [arXiv:hep-th/0307195].
}

\lref\KMSS{I.~R.~Klebanov, J.~Maldacena and N.~Seiberg, ``Unitary
and complex matrix models as 1-d type 0 strings,'' Commun.\ Math.\
Phys.\  {\bf 252}, 275 (2004) [arXiv:hep-th/0309168].
}

\lref\Dan{
U.~H.~Danielsson,
``A matrix model black hole: Act II,''
JHEP {\bf 0402}, 067 (2004)
[arXiv:hep-th/0312203].
}

\lref\GTT{
S.~Gukov, T.~Takayanagi and N.~Toumbas,
``Flux backgrounds in 2D string theory,''
JHEP {\bf 0403}, 017 (2004)
[arXiv:hep-th/0312208];
T.~Takayanagi,
``Comments on 2D type IIA string and matrix model,''
JHEP {\bf 0411}, 030 (2004)
[arXiv:hep-th/0408086].
}

\lref\KKK{ V.~Kazakov, I.~K.~Kostov and D.~Kutasov, ``A matrix
model for the two-dimensional black hole,'' Nucl.\ Phys.\ B {\bf
622}, 141 (2002) [arXiv:hep-th/0101011].
}

\lref\ST{ A.~Strominger and T.~Takayanagi, ``Correlators in
timelike bulk Liouville theory,'' Adv.\ Theor.\ Math.\ Phys.\
{\bf 7}, 369 (2003) [arXiv:hep-th/0303221].
}

\lref\SCHM{V.~Schomerus, ``Rolling tachyons from Liouville
theory,'' JHEP {\bf 0311}, 043 (2003) [arXiv:hep-th/0306026].
}

\lref\HiTa{ Y.~Hikida and T.~Takayanagi, ``On solvable
time-dependent model and rolling closed string tachyon,'' Phys.\
Rev.\ D {\bf 70}, 126013 (2004) [arXiv:hep-th/0408124].
}

\lref\witteng{ E.~Witten, ``Ground ring of two-dimensional string
theory,'' Nucl.\ Phys.\ B {\bf 373}, 187 (1992)
[arXiv:hep-th/9108004];
E.~Witten and B.~Zwiebach, ``Algebraic structures and differential
geometry in $2-D$ string theory,'' Nucl.\ Phys.\ B {\bf 377}, 55
(1992) [arXiv:hep-th/9201056].
}

\lref\IM{
N.~Itzhaki and J.~McGreevy,
``The large N harmonic oscillator as a string theory,''
Phys.\ Rev.\ D {\bf 71}, 025003 (2005)
[arXiv:hep-th/0408180].
}

\lref\TLD{
T.~Takayanagi,
``Matrix model and time-like linear dilaton matter,''
JHEP {\bf 0412}, 071 (2004)
[arXiv:hep-th/0411019].
}

\lref\OV{
H.~Ooguri and C.~Vafa,
``Two-Dimensional Black Hole and Singularities
of CY Manifolds,''
Nucl.\ Phys.\ B {\bf 463}, 55 (1996)
[arXiv:hep-th/9511164].
}

\lref\MKV{
S.~Mukhi and C.~Vafa,
``Two-dimensional black hole as a topological
coset model of c = 1 string
theory,''
Nucl.\ Phys.\ B {\bf 407}, 667 (1993)
[arXiv:hep-th/9301083].
}

\lref\HO{
A.~Hanany, Y.~Oz and M.~Ronen Plesser,
``Topological Landau-Ginzburg formulation and
integrable structure of 2-d
string theory,''
Nucl.\ Phys.\ B {\bf 425}, 150 (1994)
[arXiv:hep-th/9401030].
}


\lref\GM{
D.~Ghoshal and S.~Mukhi,
``Topological Landau-Ginzburg model
of two-dimensional string theory,''
Nucl.\ Phys.\ B {\bf 425}, 173 (1994)
[arXiv:hep-th/9312189].
}

\lref\giveon{
A.~Giveon, A.~Konechny, A.~Pakman and A.~Sever,
``Type 0 strings in a 2-d black hole,''
JHEP {\bf 0310}, 025 (2003)
[arXiv:hep-th/0309056].
}

\lref\BK{
M.~Bershadsky and D.~Kutasov,
``Comment on gauged WZW theory,''
Phys.\ Lett.\ B {\bf 266}, 345 (1991).
}

\lref\MO{
J.~M.~Maldacena and H.~Ooguri,
``Strings in AdS(3) and SL(2,R) WZW model. I,''
J.\ Math.\ Phys.\  {\bf 42}, 2929 (2001)
[arXiv:hep-th/0001053];
J.~M.~Maldacena, H.~Ooguri and J.~Son,
``Strings in AdS(3) and the SL(2,R) WZW model.
II: Euclidean black hole,''
J.\ Math.\ Phys.\  {\bf 42}, 2961 (2001)
[arXiv:hep-th/0005183];
J.~M.~Maldacena and H.~Ooguri,
``Strings in AdS(3) and the SL(2,R) WZW model.
III: Correlation  functions,''
Phys.\ Rev.\ D {\bf 65}, 106006 (2002)
[arXiv:hep-th/0111180].
}

\lref\WI{
E.~Witten,
``On The Structure Of The Topological Phase
Of Two-Dimensional Gravity,''
Nucl.\ Phys.\ B {\bf 340}, 281 (1990);
``The N matrix model and gauged WZW models,''
Nucl.\ Phys.\ B {\bf
371}, 191 (1992).
}

\lref\Wibh{
E.~Witten,
``On string theory and black holes,''
Phys.\ Rev.\ D {\bf 44}, 314 (1991).
}

\lref\MSW{G.~Mandal, A.~M.~Sengupta and S.~R.~Wadia,
``Classical solutions of two-dimensional string theory,''
Mod.\ Phys.\ Lett.\ A {\bf 6}, 1685 (1991).
}

\lref\HoKa{
K.~Hori and A.~Kapustin,
``Duality of the fermionic 2d black hole and
N = 2 Liouville theory as  mirror symmetry,''
JHEP {\bf 0108}, 045 (2001)
[arXiv:hep-th/0104202].
}

\lref\KL{
K.~Li,
``Topological Gravity With Minimal Matter,''
Nucl.\ Phys.\ B {\bf 354}, 711 (1991);
``Recursion Relations In Topological Gravity With Minimal Matter,''
Nucl.\ Phys.\ B {\bf 354}, 725 (1991).
}

\lref\KSTI{J.~L.~Karczmarek and A.~Strominger, ``Matrix
cosmology,'' JHEP {\bf 0404}, 055 (2004) [arXiv:hep-th/0309138];
}

\lref\EWK{J.~Polchinski, ``Classical limit of (1+1)-dimensional
string theory,'' Nucl.\ Phys.\ B {\bf 362}, 125 (1991);

D.~Minic, J.~Polchinski and Z.~Yang, ``Translation invariant
backgrounds in (1+1)-dimensional string theory,'' Nucl.\ Phys.\ B
{\bf 369}, 324 (1992);

G.~W.~Moore and R.~Plesser,
``Classical scattering in (1+1)-dimensional string theory,''
Phys.\ Rev.\ D {\bf 46}, 1730 (1992)
[arXiv:hep-th/9203060];

A.~Dhar, G.~Mandal and S.~R.~Wadia,
``A Time dependent classical solution of c = 1
string field theory and
Int.\ J.\ Mod.\ Phys.\ A {\bf 8} (1993) 3811
[arXiv:hep-th/9212027].
}

\lref\TDC{J.~L.~Karczmarek and A.~Strominger,
``Closed string tachyon condensation at c = 1,'' JHEP {\bf 0405},
062 (2004) [arXiv:hep-th/0403169];

S.~R.~Das, J.~L.~Davis, F.~Larsen and P.~Mukhopadhyay, ``Particle
production in matrix cosmology,'' Phys.\ Rev.\ D {\bf 70}, 044017
(2004) [arXiv:hep-th/0403275];

P.~Mukhopadhyay, ``On the problem of particle production in c = 1
matrix model,'' JHEP {\bf 0408}, 032 (2004)
[arXiv:hep-th/0406029];

S.~R.~Das and J.~L.~Karczmarek, ``Spacelike boundaries from the c
= 1 matrix model,'' arXiv:hep-th/0412093;

S.~Hirano, ``Energy quantisation in bulk bouncing tachyon,''
arXiv:hep-th/0502199;

S.~R.~Das, ``Non-trivial 2d space-times from matrices,''
arXiv:hep-th/0503002.
}

\lref\WiTSM{ E.~Witten, ``Topological Sigma Models,'' Commun.\
Math.\ Phys.\  {\bf 118}, 411 (1988).
}

\lref\GV{
D.~Ghoshal and C.~Vafa,
``C = 1 string as the topological theory of the conifold,''
Nucl.\ Phys.\ B {\bf 453}, 121 (1995)
[arXiv:hep-th/9506122].
}

\lref\KS{
Y.~Kazama and H.~Suzuki,
``New N=2 Superconformal Field Theories And
Superstring Compactification,''
Nucl.\ Phys.\ B {\bf 321}, 232 (1989).
}

\lref\EY{
T.~Eguchi and S.~K.~Yang,
``N=2 Superconformal Models As Topological Field Theories,''
Mod.\ Phys.\ Lett.\ A {\bf 5}, 1693 (1990).
}

\lref\MOK{
E.~Martinec and K.~Okuyama,
``Scattered results in 2D string theory,''
JHEP {\bf 0410}, 065 (2004)
[arXiv:hep-th/0407136].
}

\lref\MaLO{
J.~Maldacena,
``Long strings in two dimensional string theory
and non-singlets in the matrix model,''
arXiv:hep-th/0503112.
}

\lref\KMSBH{
J.~L.~Karczmarek, J.~Maldacena and A.~Strominger,
``Black hole non-formation in the matrix model,''
arXiv:hep-th/0411174.
}

\lref\FV{
J.~J.~Friess and H.~Verlinde,
``Hawking effect in 2-D string theory,''
arXiv:hep-th/0411100.
}


\lref\SY{
T.~Suyama and P.~Yi,
``A holographic view on matrix model of black hole,''
JHEP {\bf 0402}, 017 (2004)
[arXiv:hep-th/0401078].
}

\lref\ES{ T.~Eguchi and Y.~Sugawara, ``SL(2,R)/U(1) supercoset and
elliptic genera of non-compact Calabi-Yau manifolds,'' JHEP {\bf
0405}, 014 (2004) [arXiv:hep-th/0403193];
 ``Conifold type singularities, N = 2 Liouville and SL(2:R)/U(1)
theories,'' JHEP {\bf 0501}, 027 (2005) [arXiv:hep-th/0411041].
}

\lref\Dual{
H.~Dorn and H.~J.~Otto, ``Two and three point functions
in Liouville theory,'' Nucl.\ Phys.\ B {\bf 429}, 375 (1994)
[arXiv:hep-th/9403141];

A.~B.~Zamolodchikov and A.~B.~Zamolodchikov, ``Structure
constants and conformal bootstrap in Liouville field theory,''
Nucl.\ Phys.\ B {\bf 477}, 577 (1996) [arXiv:hep-th/9506136];

J.~Teschner, ``Liouville theory revisited,'' Class.\ Quant.\
Grav.\  {\bf 18}, R153 (2001) [arXiv:hep-th/0104158];

Y.~Nakayama, ``Liouville field theory: A decade after the
revolution,'' Int.\ J.\ Mod.\ Phys.\ A {\bf 19}, 2771 (2004)
[arXiv:hep-th/0402009].
}

\lref\DDK{ F.~David, ``Conformal Field Theories Coupled To 2-D
Gravity In The Conformal Gauge,'' Mod.\ Phys.\ Lett.\ A {\bf 3},
1651 (1988);

J.~Distler and H.~Kawai, ``Conformal Field Theory And 2-D Quantum
Gravity Or Who's Afraid Of Joseph Liouville?,'' Nucl.\ Phys.\ B
{\bf 321}, 509 (1989).
}

\lref\DVV{ R.~Dijkgraaf, H.~Verlinde and E.~Verlinde,
``Topological Strings In D $<$ 1,'' Nucl.\ Phys.\ B {\bf 352}, 59
(1991).
}

\lref\Ts{
J.~Teschner,
``On structure constants and fusion rules in the
SL(2,C)/SU(2) WZNW  model,''
Nucl.\ Phys.\ B {\bf 546}, 390 (1999)
[arXiv:hep-th/9712256];
``Operator product expansion and factorization
in the H-3+ WZNW model,''
Nucl.\ Phys.\ B {\bf 571}, 555 (2000)
[arXiv:hep-th/9906215].
}

\lref\SenOC{
A.~Sen,
``Open-closed duality: Lessons from matrix model,''
Mod.\ Phys.\ Lett.\ A {\bf 19}, 841 (2004)
[arXiv:hep-th/0308068].
}

\lref\GK{
A.~Giveon and D.~Kutasov,
``Little string theory in a double scaling limit,''
JHEP {\bf 9910}, 034 (1999)
[arXiv:hep-th/9909110];
``Comments on double scaled little string theory,''
JHEP {\bf 0001}, 023 (2000)
[arXiv:hep-th/9911039];
``Notes on AdS(3),'' Nucl.\ Phys.\ B {\bf 621}, 303 (2002)
[arXiv:hep-th/0106004].
}

\lref\HS{
K.~Hosomichi and Y.~Satoh,
``Operator product expansion in string theory on AdS(3),''
Mod.\ Phys.\ Lett.\ A {\bf 17}, 683 (2002)
[arXiv:hep-th/0105283];
Y.~Satoh,
``Three-point functions and operator product
expansion in the SL(2)  conformal field theory,''
Nucl.\ Phys.\ B {\bf 629}, 188 (2002)
[arXiv:hep-th/0109059].
}

\lref\GN{
G.~Giribet and C.~Nunez,
``Correlators in AdS(3) string theory,''
JHEP {\bf 0106}, 010 (2001)
[arXiv:hep-th/0105200].
}

\lref\WiTS{
E.~Witten,
``Mirror manifolds and topological field theory,''
arXiv:hep-th/9112056.
}

\lref\TTD{T.~Takayanagi and S.~Terashima,``
c=1 Matrix Model from String Field Theory,''
arXiv:hep-th/0503184.
}

\lref\RT{
S.~Ribault and J.~Teschner,
``H(3)+ correlators from Liouville theory,''
arXiv:hep-th/0502048.
}

\lref\HHS{
Y.~Hikida, K.~Hosomichi and Y.~Sugawara,
``String theory on AdS(3) as discrete light-cone Liouville theory,''
Nucl.\ Phys.\ B {\bf 589}, 134 (2000)
[arXiv:hep-th/0005065].
}

\lref\WIL{E.~Witten,``Phases of N = 2 theories in two
dimensions,'' Nucl.\ Phys.\ B {\bf 403}, 159 (1993)
[arXiv:hep-th/9301042].
}

\lref\EKYY{ T.~Eguchi, H.~Kanno, Y.~Yamada and S.~K.~Yang,
``Topological strings, flat coordinates and gravitational
descendants,'' Phys.\ Lett.\ B {\bf 305}, 235 (1993)
[arXiv:hep-th/9302048].
}

\lref\Lo{ A.~Lossev, `` Descendants constructed from matter fields
in topological Landau-Ginzburg theories coupled to topological
gravity ,'' arXiv:hep-th/9211090.
}

\lref\ADKMV{ M.~Aganagic, R.~Dijkgraaf, A.~Klemm, M.~Marino and
C.~Vafa, ``Topological strings and integrable hierarchies,''
arXiv:hep-th/0312085.
}

\lref\Kap{ A.~Kapustin, ``Noncritical superstrings in a
Ramond-Ramond background,'' JHEP {\bf 0406}, 024 (2004)
[arXiv:hep-th/0308119].
}


\lref\Wakimoto{ M.~Wakimoto, ``Fock Representations Of The Affine
Lie Algebra A1(1),'' Commun.\ Math.\ Phys.\  {\bf 104}, 605
(1986).
}

\lref\Se{ N.~Seiberg, ``Notes On Quantum Liouville Theory And
Quantum Gravity,'' Prog.\ Theor.\ Phys.\ Suppl.\  {\bf 102}, 319
(1990).
}

\lref\LVW{ W.~Lerche, C.~Vafa and N.~P.~Warner, ``Chiral Rings In
N=2 Superconformal Theories,'' Nucl.\ Phys.\ B {\bf 324}, 427
(1989).
}

\lref\LZ{ B.~H.~Lian and G.~J.~Zuckerman, ``New selection rules
and physical states in 2-D gravity: Conformal gauge,'' Phys.\
Lett.\ B {\bf 254}, 417 (1991);
``2-D gravity with c = 1 matter,'' Phys.\ Lett.\ B {\bf 266}, 21
(1991).
}

\lref\Xi{ X.~Yin, ``Matrix models, integrable structures, and
T-duality of type 0 string theory,'' arXiv:hep-th/0312236.
}

\lref\PaSu{ J.~Park and T.~Suyama, ``Type 0A matrix model of black
hole, integrability and holography,'' arXiv:hep-th/0411006.
}

\lref\GRT{ D.~Gaiotto, L.~Rastelli and T.~Takayanagi, ``Minimal
superstrings and loop gas models,'' arXiv:hep-th/0410121.
}

\lref\HPT{ A.~Hanany, N.~Prezas and J.~Troost, ``The partition
function of the two-dimensional black hole conformal  field
theory,'' JHEP {\bf 0204}, 014 (2002) [arXiv:hep-th/0202129].
}

\lref\OS{
 H.~Ita, H.~Nieder, Y.~Oz and T.~Sakai,
 ``Topological B-model, matrix models,
c-hat = 1 strings and quiver gauge
theories,'' JHEP {\bf 0405}, 058 (2004) [arXiv:hep-th/0403256];

U.~H.~Danielsson, M.~E.~Olsson and M.~Vonk, ``Matrix models, 4D
black holes and topological strings on non-compact Calabi-Yau
manifolds,'' JHEP {\bf 0411}, 007 (2004) [arXiv:hep-th/0410141];

S.~Hyun, K.~Oh, J.~D.~Park and S.~H.~Yi, ``Topological B-model and
c-hat = 1 string theory,'' arXiv:hep-th/0502075.
}

\lref\OOV{ H.~Ooguri and C.~Vafa, ``Worldsheet derivation of a
large N duality,'' Nucl.\ Phys.\ B {\bf 641}, 3 (2002)
[arXiv:hep-th/0205297];

T.~Okuda and H.~Ooguri, ``D branes and phases on string
worldsheet,'' Nucl.\ Phys.\ B {\bf 699}, 135 (2004)
[arXiv:hep-th/0404101].
}

\lref\matreview{I.~R.~Klebanov, ``String theory in
two-dimensions,'' arXiv:hep-th/9108019;

P.~H.~Ginsparg and G.~W.~Moore, ``Lectures on 2-D gravity and 2-D
string theory,'' arXiv:hep-th/9304011;

J.~Polchinski, ``What is string theory?,'' arXiv:hep-th/9411028.
}

\lref\OH{N.~Ohta and H.~Suzuki, ``Bosonization of a topological
coset model and noncritical string theory,'' Mod.\ Phys.\ Lett.\ A
{\bf 9}, 541 (1994) [arXiv:hep-th/9310180].
}

\lref\OHT{K.~Itoh, H.~Kunitomo, N.~Ohta and M.~Sakaguchi, ``BRST
Analysis of physical states in two-dimensional black hole,''
Phys.\ Rev.\ D {\bf 48}, 3793 (1993) [arXiv:hep-th/9305179].
}

\lref\AH{ O.~Aharony, O.~Ganor, J.~Sonnenschein, S.~Yankielowicz
and N.~Sochen, ``Physical states in G/G models and 2-d gravity,''
Nucl.\ Phys.\ B {\bf 399}, 527 (1993) [arXiv:hep-th/9204095];
O.~Aharony, O.~Ganor, J.~Sonnenschein and S.~Yankielowicz, ``c = 1
string theory as a topological G/G model,'' Phys.\ Lett.\ B {\bf
305}, 35 (1993) [arXiv:hep-th/9302027].
}

\lref\Nak{Y.~Nakayama, ``Crosscap states in N = 2 Liouville
theory,'' Nucl.\ Phys.\ B {\bf 708}, 345 (2005)
[arXiv:hep-th/0409039].
}

\lref\Gr{ G.~E.~Giribet and D.~E.~Lopez-Fogliani, ``Remarks on
free field realization of SL(2,R)k/U(1) x U(1) WZNW model,'' JHEP
{\bf 0406}, 026 (2004) [arXiv:hep-th/0404231];
K.~Hosomichi, ``N = 2 Liouville theory with boundary,''
arXiv:hep-th/0408172.
}

\lref\HIV{
K.~Hori, A.~Iqbal and C.~Vafa,
``D-branes and mirror symmetry,''
arXiv:hep-th/0005247.
}

\baselineskip 18pt plus 2pt minus 2pt

\Title{\vbox{\baselineskip12pt \hbox{hep-th/0503237}
\hbox{HUTP-05/A0015}
  }}
{\vbox{\centerline{$c<1$ String from Two Dimensional Black Holes}}}
\centerline{Tadashi Takayanagi\foot{e-mail:
takayana@bose.harvard.edu}}

\medskip\centerline{ \it Jefferson Physical Laboratory}
\centerline{\it Harvard University}
\centerline{\it Cambridge, MA 02138, USA}

\vskip .5in \centerline{\bf Abstract} We study a topological
string description of
the $c<1$ non-critical string whose matter part is
defined by the time-like linear dilaton CFT. We show that the
topologically twisted $N=2$ $SL(2,R)/U(1)$ model (or
supersymmetric 2D black hole) is equivalent to the $c<1$
non-critical string compactified at a specific radius by comparing
their physical spectra and correlation functions. We
examine another equivalent description in the topological
Landau-Ginzburg model and check that it reproduces the same
scattering amplitudes. We also discuss its matrix model dual
description.

\noblackbox

\Date{April, 2005}

\listtoc
\writetoc

\newsec{Introduction}

The two dimensional (2d) string theory has been a very useful
laboratory of quantum gravity. This is because it can be exactly
solvable by the dual $c=1$ matrix model description \tdmat
\matreview\ including all loop corrections. Recently, remarkable
progresses have been made such as the holographic interpretation
of the dual matrix model \MV \KMS \SenOC \TTD, the
non-perturbatively stable construction of the type 0 matrix model
\TT \six \KMSS, the applications to time-dependent
backgrounds\foot{Early relevant works can be found in \EWK.} \AKKT
\KSTI \TDC \TLD\ and so on. Clearly, these ideas are very helpful
when we would like to understand the dynamical properties of the
2d string theory as a simplest example of quantum gravity.

One of the most important unsolved problems in 2d string theory is
a clear understanding of the 2d black hole \Wibh \MSW. This
background has been known to be described by the $SL(2,R)/U(1)$
coset CFT \Wibh\ and its matrix model dual was proposed \KKK\ in
the Euclidean case. Nevertheless, many of its physical properties
of the black holes such as its entropy etc. have been poorly
understood even at present\foot{For recent progresses on this
subject refer to e.g. \giveon \Dan \GTT \Xi \SY \MOK \PaSu \FV \KMSBH
\MaLO.}. One interesting way to make an important progress in such
a difficult problem in string theory is to direct our attention to
its topological properties. For example, when we are interested
in the critical superstring compactified on a Calabi-Yau space, we
can extract important holomorphic quantities by considering the
topological string \WiTSM \WiTS\ on that space. In the same way,
we can expect that the topologically twisted $N=2$ $SL(2,R)/U(1)$
coset model (Kazama-Suzuki model \KS) captures certain significant
properties of 2d black holes in superstring \KKK \HoKa \giveon. In
this case, we have an advantage that the topological twisting does
not reduce the degree of freedom so much. This is because the 2d
string theory itself does not include infinitely many massive
fields as opposed to the ordinary ten dimensional superstring.
Indeed, it has been known that such kinds of lower dimensional
string theories (i.e. non-critical strings) often possess
equivalent descriptions in terms of topological strings \ADKMV.

Such an example, which is the most relevant to us, will be the
well-known equivalence \WI \MKV\ between the twisted $N=2$
$SL(2,R)_{3}/U(1)$ coset model at the level $k=3$, and the $c=1$
string (or equally the 2d string) compactified at the self-dual
radius. This twisted theory can be geometrically interpreted as
the topological string on the conifold \GV \OV. It is also known
to be equivalent\foot{This equivalence is essentially the same as
the mirror symmetry which relates the $N=2$ Liouville theory with
the $N=2$ $SL(2,R)/U(1)$ model \GK \KKK \HoKa \ES.} to the
topological Landau-Ginzburg (LG) model defined by the
superpotential $W=X^{-1}$ \GM \HO \OV \OOV. However, in this
example its relation to the physical type 0 string on the 2d black
hole is not so clear since its critical condition requires the
different value $k=3/2$ of the level.

Motivated by this, we would like to discuss the topologically
twisted $N=2$ $SL(2,R)_{k}/U(1)$ model at general values of the
level $k$ in the present paper. We will argue that it is
equivalent to a specific compactified $c<1$ string whose matter
part is defined by the time-like linear dilaton CFT \TLD\ instead
of minimal models. We will often call this theory a non-minimal
$c<1$ string below just for simplicity. The results in this paper
are conveniently summarized in the final section. If we look at
this equivalence reversely, it also reveals the essential
topological properties of the $c<1$ string theory.

In this $c<1$ string, the matter CFT is described by a time-like
boson $X_0$ whose linear dilaton gradient is $q=\s{2}(1/b-b)$.
Thus the central charge of this matter CFT is $c_X=1-6q^2$. The
Liouville sector is defined in a standard way by the space-like
scalar $\phi$ whose linear dilaton gradient is $Q=\s{2}(b+1/b)$
(central charge $c_\phi=1+6Q^2$). We also assume the usual
Liouville term $\mu \int dz^2 e^{\s{2}b\phi}$. Totally this model
defines a critical bosonic string because $c_X+c_\phi=26$. After
we perform a Lorentz boost so that the coupling constant $g_s$
does not depend on time, it represents a simple time-dependent
background in 2d string theory with a non-standard Liouville
potential $\mu \int dz^2 e^{[(b^2-1)X^0+(1+b^2)\phi]/\s{2}}$ \TLD.
Indeed the matrix model dual of this background is given by the
following time-dependent fermi surface \TLD\ of $c=1$ matrix model
in the phase space $(x,p)$
\eqn\fermiss{\left(\f{-p-x}{2}\right)^{b^2}\left(\f{p-x}{2}\right)
=\mu e^{(b^2-1)t},} where $t$ is the time in the matrix model. The
special property of this model compared with other time-dependent
ones is that it is solvable even in the world-sheet theory.

The deformation parameter $b$ in the $c<1$ string is identified
with the one in the twisted model via the relation $n=b^{-2}$ (or
equally $n=b^2$) under the equivalence. We also see that these
have another equivalent description by the topological LG models
defined by the potential $W=X^{-n}$. In a particular limit $b=1$
we reproduce the known equivalence between $c=1$ string and the
topological models \WI \MKV \AH \GM \HO. The possibility of the
equivalence between the twisted $SL(2,R)/U(1)$ model and a certain
$(p,q)$ non-critical string was already mentioned in \AH \OH \OV\
when $n$ is a rational number $n=p/q$.
 In the present paper, we explicitly identify the relevant
string theory with the compactified non-minimal $c<1$ string
defined just before. This string theory can also be defined by the
dual matrix model description \TLD. In addition, we also find that
most of our results do not depend on whether $n$ is rational or
irrational except the periodicity in the Euclidean time direction.

Another aim of this paper is to check this kind of equivalence
between the topological strings and the non-critical strings
including non-trivial interactions in addition to the comparison
of the physical spectrum. One way to compute the interactions is
to examine the correlators by employing the mathematical
structures of the topological gauged WZW model \WI. It is also
possible to do this by using the equivalent description of the
topological LG models  \GM \HO. Even though these analyses
reproduce the essential part of the scattering amplitudes in the
non-critical string, the derivation of 
the non-local transformation (or almost
equally so called leg factor \matreview) for each vertex operators
is not straightforward. However, now we
can also directly compute the correlators in terms of those in the
untwisted theory owing to the recent progresses of the
understanding of $SL(2,R)$ WZW model (refer to \Ts \MO\ and
references therein). This has not been done even for the familiar
$c=1$ case \MKV. In the present paper we will explicitly compute
the three point functions in the twisted coset model and check
that they agree with those in the $c<1$ string including the $c=1$
case as a particular limit.

The organization of this paper is as follows. In section two, we
examine the twisted $SL(2,R)/U(1)$ model. We find its physical
states in the free field representation and show that it is the
same as those in the $c<1$ string. In section three, we directly
compute the three point function in the twisted coset model and
check that they agree with those in the $c<1$ string. In section
four, we discuss the topological LG model, which is expected to be
equivalent to the twisted $SL(2,R)/U(1)$ model. Indeed we check
the model reproduces the correct scattering amplitudes of tachyons
in the $c<1$ string. In section five we summarize the conclusions
and discuss future problems.

\newsec{Twisting 2D Black holes}

We would like to investigate the topological (A-model) twist of
the $N=2$ coset model $\f{SL(2,R)}{U(1)}$. This SCFT describes the
supersymmetric version of 2d black hole \Wibh \MSW \KKK. What we
would like to show is that the spectrum of physical states agrees
with that of the non-minimal $c<1$ string \TLD\ whose matter part
is defined by the time-like linear dilaton theory. Since the
argument here is a natural extension of those in the $c=1$ string
case \MKV, the discussions below will be a bit brief and we will
share almost the same notations. Nevertheless, we will also add
several clarifications in the light of the modern understandings
of the $SL(2,R)$ WZW model (see \OV\ and references therein). We
 use the normalization of OPEs in the $\al=2$ unit in the most
part of this paper.

\subsec{Description of Coset Model}

The $N=2$ $\f{SL(2,R)}{U(1)}$ coset model\foot{We can also write
the supersymmetric model as $\f{SL(2,R)_{n+2}\times U(1)}{U(1)}$.
The $U(1)$ part in the numerator corresponds to the fermions and
we write this by $\psi$ and $\bar{\psi}$.} at level $n(>0)$ is
equivalent to the product of the bosonic $SL(2,R)_{n+2}$ WZW model
(at level $k=n+2$) and a Dirac free fermion $(\psi,\bar{\psi})$.
In this section we do not have to assume that $n$ is an integer.
Whether it is an integer or not becomes relevant when we consider
its interpretation as a non-critical string theory. We will
discuss this issue in section 5.2.

We employ the Wakimoto free field representation \Wakimoto\ for
the bosonic $SL(2,R)_{n+2}$ WZW model via the bosonization\foot{
In the most part of this paper we make explicit only the
left-moving (or chiral) sector. The right-moving (anti-chiral)
sector can be constructed in the same way. The world-sheet
coordinate is denoted by $(z,\bar{z})$.} \eqn\slt{J^{-}=\beta,\
\ \ J^{+}=\beta\gamma^2-\s{2n}\gamma\de\phi +(n+2)\de\gamma,\ \ \
J^3=\beta\gamma-\s{\f{n}{2}}\de\phi.} The OPEs are defined by
$\phi(z)\phi(0)\sim -\log z,\ \ \ \beta(z)\gamma(0)\sim 1/z$. The
operators \slt\ satisfy the $SL(2,R)_{n+2}$ current algebra
\eqn\opecu{J^3(z)J^3(0)\sim -\f{(n+2)/2}{z^2},\ \ \
J^3(z)J^{\pm}(0)\sim \pm \f{J^{\pm}(z)}{z^2},\ \ \
J^{+}(z)J^{-}(0)\sim \f{n+2}{z^2}-\f{2J^3(0)}{z}.}
The stress energy tensor is given by
\eqn\str{T=\beta\de\gamma-\f{1}{2}(\de\phi)^2
+\f{1}{\s{2n}}\de^2\phi.}

The primary fields $\Phi_{j,m}$ in the $SL(2,R)$ current algebra
are expressed as ($m$ is the eigenvalue of $J^3_0$)
\eqn\prslt{\Phi_{j,m}=\gamma^{j+m}e^{-\s{\f{2}{n}}j\phi}.} Then we
can construct the lowest weight discrete representation $D^+_{j}:
m=-j,-j+1,-j+2,\ddd$ starting from the lowest weight state (LWS)
$\Phi_{j,-j}=e^{-\s{\f{2}{n}}j\phi}$. We can also find its
conjugate representation by identifying its LWS with
\eqn\prslll{\Phi_{j,-j}'=(\beta)^{n+2j+1}
e^{\s{\f{2}{n}}(j+n+1)\phi}.} On the other hand, the highest
weight discrete representation is denoted by
$D^-_{j}:m=j,j-1,j-2,\ddd$. Its highest weight state (HWS) has the
spin $m=j$. Since the background
charge\foot{We define the background charge for a
field $\phi$ by $T=-\f{1}{2}(\de\phi)^2+\f{Q}{2}\de^2 \phi$. The
string coupling constant behaves like $g_s=e^{\f{Q}{2}\phi}$. The
conformal dimension of the primary field $e^{\alpha\phi}$ is given
by $\Delta(e^{\alpha\phi})=\f{1}{2}\ap(Q-\ap)$.}
 of $\phi$ is $Q(\phi)=\s{\f{2}{n}}$ as can be seen from \str,
we can check that the conformal dimension of the
operators $\Phi_{j,m}$ and $\Phi'_{j,m}$ is $\Delta=-\f{j(j+1)}{n}$.

In order to
define the coset, we can gauge the following $U(1)$ current
(see e.g. \BK \OV) \eqn\uo{J_g=J^3-\bar{\psi}\psi-i\s{\f{n}{2}}
\de X=\hat{J}^3-i\s{\f{n}{2}}\de X,} where $\hat{J}^3$ is the
third current of the super $SL(2,R)_n$ $N=1$ WZW model. Note also the
OPEs $X(z)X(0)\sim -\log(z)$ and $\psi(z)\bar{\psi}(0)\sim
\f{1}{z}$. To perform the $U(1)$ quotient we can add the $c=-2$
ghosts\foot{Their conformal dimensions are $\Delta(\xi)=0$ and
$\Delta(\eta)=1$.} $(\xi,\eta)$ and define the BRST charge
$Q_B=\int dz \xi(z) J_g(z)$. The $U(1)$ current of the $N=2$ SCFT
is \eqn\crr{J_{R}=\f{n+2}{n}\bar{\psi}\psi-\f{2}{n}J^3\simeq
\bar{\psi}\psi-i\s{\f{2}{n}}\de X.} Notice that $J_q(z)J_R(0)\sim
0$. Then we can equivalently use the following $U(1)$ current of
the $N=2$ SCFT (see also \OV)
\eqn\mocrr{J'_{R}=J_{R}-2J_g=3\bar{\psi}\psi-2J^3
+i\s{\f{2}{n}}(n-1)\de
X.} Geometrically the quotient $SL(2,R)/U(1)$ looks like a cigar
(or 2d black hole) with the asymptotic radius $R=\s{2n}$.

\subsec{Topological Twist}

Now let us take the topological (A-model\foot{Here we
define the topological string theory by the A-model twist i.e.
$T(z)\to T(z)+\f{1}{2}\de J(z)$ and $T(\bar{z})\to
T(\bar{z})-\f{1}{2}\bar{\de} J(\bar{z})$ \WiTS \EY. The
anti-chiral (or right-moving) counterpart of the R-current \mocrr\
is given by
$J'_{R}(\bar{z})=3\bar{\psi}\psi+2J^3
+i\s{\f{2}{n}}(n-1)\bar{\de} X$ in
our convention; the right-moving gauging current is $J_g(\bar{z})
=J^3(\bar{z})+3\psi\bar{\psi}+i\s{\f{2}{n}}\bar{\de} X$. Thus the
twist acts on the $SL(2,R)$ part symmetrically, while it does on
the $U(1)$ boson $X$ asymmetrically. Notice that we do not want to
perform the B-model twist because in that case the background
charge for $\phi$ in the left-moving section takes the sign
opposite to the one in the right-moving sector.})
twist \WiTS\ of the $N=2$
coset via the standard rule $T\to T+\f{1}{2}\de J$ \EY. The
background charges
 become $Q'({\phi})=\s{\f{2}{n}}+\s{2n}$ and
$Q'(X)=i(\s{2n}-\s{\f{2}{n}})$. The central charges become
$c=1+6\f{(1+n)^2}{n}$ and $c=1-\f{6(n-1)^2}{n}$, respectively. In
summary we started with the fields ($\Delta(A)$ denotes the
conformal dimension of the operator $A$) \eqn\summa{\eqalign{ X:&
\ \  c=1,\ \  Q(X)=0 \cr \phi :& \ \ c=1+6/n,\ \  Q(\phi)=
\s{\f{2}{n}} \cr (\psi,\bar{\psi}) :& \ \ c=1, \ \
\Delta(\psi)=\Delta(\bar{\psi})=1/2 \cr (\beta,\gamma) :& \ \ c=2,
\ \ \Delta(\beta)=1, \ \Delta(\gamma)=0 \cr (\eta,\xi) :& \ \
c=-2, \ \ \Delta(\eta)=1, \ \Delta(\xi)=0 ,}} and after the twist
we get \eqn\summa{\eqalign{ X:& \ \ c=1-6(n-1)^2/n,\ \
Q(X)=i(\s{2n}-\s{\f{2}{n}}) \cr \phi :& \ \ c=1+6(1+n)^2/n,\ \
Q(\phi)= \s{\f{2}{n}}+\s{2n} \cr (\psi,\bar{\psi}) :& \ \ c=-26, \
\ \Delta(\psi)=2,\ \ \Delta(\bar{\psi})=-1 \cr (\beta,\gamma) :& \
\ c=2, \ \ \Delta(\beta)=0, \ \Delta(\gamma)=1 \cr (\eta,\xi) :& \
\ c=-2, \ \ \Delta(\eta)=1, \ \Delta(\xi)=0 .}} Notice that the
total central charge after the twist is zero as expected.

Then we can interpret the twisted system as a critical bosonic
string. The boson $X$ is a free boson with the linear dilaton and
$\phi$ is the Liouville field on the world-sheet. In the 2d
spacetime viewpoint, $X$ and $\phi$ are the (Euclidean) time and
space coordinate. The fermions $(\psi, \bar{\psi})$ correspond to
the $(b,c)$ ghosts. The fields $(\beta,\gamma)$ and $(\eta,\xi)$
are almost canceled with each other. The original screening term
in the $SL(2,R)$ WZW model \eqn\scr{\int
dz^2\beta(z)\ti{\beta}(\bar{z})e^{\s{\f{2}{n}}\phi},} can be
regarded as the Liouville potential term in the $c<1$ string
\eqn\scrr{\mu \int dz^2 e^{\s{\f{2}{n}}\phi},} since $\beta$
becomes conformal dimension zero after the twist and can be treated
as a constant. On the other
hand, there is no screening term for the $X$ field. Thus this
string theory is equivalent to the non-minimal $c<1$ string \TLD\
(also this is reviewed in the introduction) at
$b=\f{1}{\s{n}}$ (or equally $b=\s{n}$).
 Notice also the Euclidean
`time' $X$ is compactified and its radius\foot{Below we also
consider the $Z_n$ orbifold of the coset $SL(2,R)/U(1)$. In that
case we have the different radius $R=\s{2/n}$.} is again given by
$R=\s{2n}$.

\subsec{Chiral Primaries}

Here we would like to find the chiral primaries in the coset $N=2$
SCFT because they are obvious candidates for physical states in the
topologically twisted theory. Before we go on, we define several
free bosons to make notations simple\foot{We will closely follow
the notations in \giveon. However, the definition of $J_R$ has
opposite sign to \giveon.}. First let us bosonize the fermion as
\eqn\bosniz{\psi(z)=e^{iH(z)},\ \ \ \bar{\psi}(z)= e^{-iH(z)},\ \
\psi(z)\bar{\psi}(z)=i\de H(z).} We also define bosons $X_3$ and
$X_{R}$ from $J^3(z)$ and $J_{R}(z)$ as follows
\eqn\defbos{J^3(z)=-\s{\f{n+2}{2}}\de X_3(z),\ \ \ \
J_{R}(z)=-i\s{\f{n+2}{n}}\de X_R(z).} We can also rewrite \crr\ as
follows
\eqn\relationg{X_{R}=\s{\f{n+2}{n}}H+i\s{\f{2}{n}}X_3.} These
bosons are normalized such that their OPEs are given by
$H(z)H(0)\sim -\log (z)$, $X_3(z)X_3(0)\sim -\log (z)$ and
$X_{R}(z)X_{R}(0)\sim -\log (z).$

The primary fields $\Phi_{j,m}$
in the bosonic $SL(2,R)$ WZW can be expressed
as \eqn\primarysl{\Phi_{j,m}=V_{jm}e^{\s{\f{2}{n+2}}mX_3},}
where $V_{jm}$ is the primary of $SL(2,R)/U(1)$ coset.
The conformal dimensions are
$\Delta(\Phi_{j,m})=-\f{j(j+1)}{n}$.
$\Delta(V_{jm})=-\f{j(j+1)}{n}+\f{m^2}{n+2}$.
Notice that the $V_{jm}$ part has no $J^3_0$ charge.

Let us return to the formulation with ghosts $(\xi,\eta)$. Because
of the BRST invariance\foot{Note that this BRST operator is
completely different from the one which defines the topological
twisted theory.} $Q_{B}=\int dz \xi(z)J_g(z)\sim 0$ on the
physical states (see \uo ), the primary operators
\eqn\primary{e^{isH}V_{jm}e^{\s{\f{2}{n+2}}m X_3},} in $N=1$
$SL(2,R)$ model, which has the $J_g$ charge $q_g=s+m$ and the
dimension $\Delta=-\f{j(j+1)}{n}+\f{s^2}{2}$, are now
dressed by $X$ as follows \eqn\drepr{
e^{isH}V_{jm}e^{\s{\f{2}{n+2}}m X_3}\cdot
e^{i\s{\f{2}{n}}(s+m)X}.} The role of $X$ dressing is that it
annihilates with the unwanted $U(1)$ part in N=1 $SL(2,R)$
operator \primary . We can find their conformal dimensions
\eqn\confgg{\Delta=\left(-\f{j(j+1)}{n}
+\f{s^2}{2}\right)+\f{(s+m)^2}{n}.}

Chiral primary states in the NS-sector of the $N=2$
coset SCFT satisfy
\eqn\chiralp{G^+_{-1/2}|NSc\lb =0.} The $N=2$
superconformal generators in the coset model \KS\ are given by
\eqn\scgene{G^+(z)=\s{\f{2}{n}}\bar{\psi}(z)J^{+}(z),\ \ \
G^-(z)=\s{\f{2}{n}}\psi(z)J^{-}(z).} We denote the operator which
corresponds to the state $|NSc\lb$ by $O_{NSc}$. Requiring the
condition \chiralp\ or $G^{+}(z)O_{NSc}(0)\sim \f{0}{z}$, we can
find the following chiral primaries (i.e. $s=0$ and $j=m$)
\eqn\cpp{ O_{NSc}=V_{jj}e^{\s{\f{2}{n+2}}j X_3}\cdot
e^{i\s{\f{2}{n}}j X}.} This has the conformal weight $\Delta=-j/n$
and R-charge $q_R=-2j/n$ (in the same way we can construct
anti-chiral state defined by $s=0$ and $j=-m$ with
$\Delta=-q_R/2=-j/n$).

We would like to perform the spectral flow\foot{Note the formula
$\Delta'=\Delta+\theta q_R+c\theta^2/6,\ \ q'_R=q_R+c\theta/3$ for
the spectral flow by the angle $\theta$ \LVW. This corresponds to
the multiplication of the operator $e^{-i\theta\s{\f{n+2}{n}}X}$.
In particular we are interested in the spectral flow from NS to
R-sector is $\theta=-1/2$. Also distinguish this spectral flow
from the one we perform in the next subsection.}
 so that we get
R-sector states $|Rc\lb$ \LVW\ that satisfy
\eqn\satg{G^+_0|Rc\lb=G^-_0|Rc\lb=0.} Notice that the first
condition shows that this is the physical state in the topological
twisted model which we are interested in.

The spectral flow corresponds to the shift of $X_R$ momentum by
$e^{\f{i}{2}\s{\f{n+2}{n}}X_R}\sim e^{\f{i}{2}H+\f{i}{\s{2n}}X}$.
Thus we find the corresponding operators in R-sector
\eqn\srtsd{V_{j,j}e^{\s{\f{2}{n+2}}j X_3}\cdot
e^{\f{i}{2}H}e^{i\s{\f{2}{n}}(j+1/2)X}.} These have the property
$\Delta=\f{n+2}{8n}$ and $q_R=-\f{2j}{n}-\f{n+2}{2n}$. Note that
the second condition in \satg\ is too restrictive to find the
physical state and there are indeed other states as we will explain
later. It is also a useful fact that the non-triviality of
cohomology with respect to $G_0^+$ requires the condition
$\Delta=\hat{c}/8=\f{n+2}{8n}$.

\subsec{Physical States}

To find the physical states in the topologically twisted theory,
we can start with the Ramond states which satisfy $G_0^+ |Rc\lb=0$
in the untwisted model and perform the spectral flow \LVW \MKV\
into the topologically twisted one (i.e. the states in NS-sector).

The physical states corresponding to the R-states \srtsd\ can be
found as follows. Because of the twisted background charges of
various bosons, the spectral flow based on the R-current
\mocrr\ is defined by
\eqn\shiftt{ |0\lb_r= e^{-\f{i}{2}\s{\f{n+2}{n}}X_R}|0\lb_t \sim
e^{-\f{3}{2}i H}
e^{i(\s{\f{n}{2}}-\f{1}{\s{2n}})X}e^{\s{\f{n+2}{2}}X_3}|0\lb_t,} ,
where $|0\lb_r$ is the Ramond sector vacuum before the twisting, and
$|0\lb_t$ is the vacuum of the topological theory. From this
argument it is obvious that the total expressions of the physical
states will include the factors \eqn\factor{e^{-iH}\cdot
e^{\s{\f{2}{n}}\left[i(j+\f{n}{2})X \right]}.} To find the result
for the bosonic $SL(2,R)$ WZW part, notice that the procedure
\shiftt\ is just the same as the $w=1$ spectral flow of $SL(2,R)$
WZW model \MO\ (see the appendix A for a brief review of the
spectral flow\foot{Refer also to \HHS\ for the spectral flow in the
free field representation.}). Since we started with the HWSs
(highest weight states) $\Phi_{j,j}$ in $D^-_{j}$, after the
spectral flow it becomes the LWSs (lowest weight states) in
$D^{-,w=1}_{j}$. Because $D^{-,w=1}_{j}$ are equivalent to
$D^+_{-j-\f{n+2}{2}}$ \MO, the states after the spectral flow can
be written as \eqn\wakilw{ \Phi_{-j-\f{n+2}{2},j+\f{n+2}{2}}=
e^{\s{\f{2}{n}}(j+\f{n+2}{2})\phi},} in the Wakimoto
representation \prslt.

These arguments lead to the following physical states in the
topological model \eqn\opb{V_j=c~
e^{\s{\f{2}{n}}\left[i(j+\f{n}{2})X+(j+\f{n+2}{2})\phi\right]},}
where we have used the fact that $\bar{\psi}=e^{-iH}$ can be
identified with the ghost $c$ after the twist \summa. Indeed these
operators have the vanishing conformal dimension. We can also
rewrite them in the following way so that they look the same as
the tachyon vertex operators in the $c<1$ string \eqn\opb{V_j=c~
e^{\f{Q(X)}{2} X+\f{Q(\phi)}{2}\phi}\cdot
e^{\s{\f{2}{n}}(j+\f{1}{2})(iX+\phi)}.} The quantum number $j$ in
our model\foot{In the standard discussions of the $SL(2,R)$ WZW
model or its $U(1)$ coset, we usually assume the unitarity bound
\MO \GK \HoKa \HPT \ES. This is given by in our convention
$-(n+1)/2<j<-1/2$. However, in this paper we do not require this
condition, so as to make the physical spectrum rich enough for our
purpose. The similar treatment seems to be valid even before the
topological twisting. One way to understand this claim intuitively
is to consider the 2d black hole ($SL(2,R)/U(1)$ model) as a
background in 2d string and try to find its physical spectrum (see
e.g.\OHT) of tachyon field. It should match with the conventional
two dimensional string in the asymptotic region $\phi\to -\infty$.
Thus we need to consider the same theory without the condition
since there are no bound for the momentum of the tachyon field in
the 2d string (see also \giveon). Notice also that it is possible
to neglect the unitarity bound in the large $n$ limit (i.e. the
classical limit) without any such assumptions since
 the condition $j<-1/2$ means the standard
non-normalizability or the Seiberg bound \Se.} takes all half
integers ($2j \in Z$) in the left-right symmetric sector (momentum
modes), while in the asymmetric sector (winding modes) we have the
different condition $2j\in nZ$. This means that the radius of $X$
is $R_A=\s{2n}$. It is also useful to consider a $Z_n$ orbifolded
model, where its $Z_n$ action acts as the shift $X\to
X+2\pi\s{2/n}$. In this case the radius is $R_A=\s{2/n}$ in the
$\al=2$ unit.

Then it is clear that the operator \opb\ can be regarded as the
tachyon vertex with the momentum $p_X=\s{\f{2}{n}}(j+1/2)$ in the
non-minimal $c<1$ string compactified at the radius
 $R=\s{2/n}$ after we take T-duality\foot{Here we took the T-duality
so that we have the same linear-dilaton gradient in both left and
right-moving direction. Notice that the A-twist leads to a
left-right asymmetric extra linear dilaton.} in the $X$ direction.
If we consider the $Z_n$ orbifolded model, then
 we get the radius $R=\s{2n}$ for the $c<1$ string.

\medskip
\medskip
\medskip

It is possible to find other physical states by acting $SL(2,R)$
currents almost in the same way as in the $n=1$ case \MKV. For $j\leq
-\f{n+2}{2}$, we can act $(J^+_{-1})^{-2j-1-n}\sim
(\gamma_{-1})^{-2j-1-n}$ on $V_j$ such that $\Delta=0$ and obtain
the following physical states \eqn\bj{B_j=
c\gamma^{-2j-n-1}e^{-i\s{\f{2}{n}}(j+\f{n+2}{2})X}
e^{\s{\f{2}{n}}(j+\f{n+2}{2})\phi}.} We can also find similar
physical states\foot{ Notice that $B_j'$ is the HWS
$\Phi_{-j-\f{n+2}{2},-j-\f{n+2}{2}}$ in
$D^{-}_{-j-\f{n+2}{2}}(=D^{+,w=-1}_{j})$.} by acting
$G^-_0$ \eqn\bjj{B_j'
=\gamma^{-2j-n-2}e^{-i\s{\f{2}{n}}(j+\f{n+2}{2})X}
e^{\s{\f{2}{n}}(j+\f{n+2}{2})\phi}.}

For $j\geq 0$ we can act
 $(J^-_0)^{2j+1}=(\beta_0)^{2j+1}$
such that $\Delta=0$ and find the physical states
\eqn\vtj{\ti{V}_j=c\beta^{2j+1}
e^{-i\s{\f{2}{n}}(j-\f{n}{2}+1)X}
e^{\s{\f{2}{n}}(j+\f{n+2}{2})\phi}.}
This is equivalent to the conjugate LWS operators
$\Phi'_{j-n/2,-j+n/2}$ (see \prslll) in
$D^{+}_{j-n/2}(=D^{-,w=1}_{-j-1})$. To interpret the operators
\vtj\ as tachyon vertexes in the $c<1$ string, we can neglect the
power of $\beta$. This is possible since $\Delta(\beta)=0$ as in
the arguments around \scr\ and \scrr. Then they looks like
\eqn\vtjtak{\ti{V}_j\sim c\ e^{\f{Q_X}{2}
X+\f{Q_{\phi}}{2}\phi}\cdot
e^{\s{\f{2}{n}}(j+\f{1}{2})(-iX+\phi)}.} Thus it is obvious from
\opb\ and \vtjtak\ that the tachyon operators $\ti{V}_j$ are dual
to $V_{-j-1}$ due to the reflection at the Liouville wall; i.e.
$\ti{V}_j\sim V_{-j-1}$.
 This relation can also be independently
seen from the $SL(2,R)$ model side by noting that $\ti{V}_j$ and
$V_{j}$ are regarded as the primary fields $\Phi_{j,j}$ and
$\Phi_{j,-j-1}$ in $D^-_j$ in the Ramond state description like
\srtsd. The operators $\Phi_{j,m}$ are equivalent
to $\Phi_{-j-1,m}$ via
the reflection relation \Ts \GK \eqn\eqivr{ \Phi_{j,m}\simeq
R(j,m,m)\Phi_{-j-1,m}.} The coefficient $R(j,m,m)$ will be
explicitly given in the section 3.1.

 In this way we have
shown that the tachyon states at the ghost number one (i.e.
$Y^{\pm}_{s,\pm s}$ in the conventional notation \witteng) can
all\foot{To be correct, there is the bound $j\geq 0$ for
$\ti{V}_{j}$. The other states with $j<0$ (i.e. $Y^+_{s,s}$) can
be obtained by acting $G^-_0$ on $\ti{T}_{j+1}$ (for the
definition of $\ti{T}_{j}$ see the arguments below).}
 be
obtained from the physical states $V_j$ and $\ti{V}_{j}$
in the twisted
$SL(2,R)/U(1)$ model. It is also possible to see
that the operators $B_j$
 and $B_j'$ can be understood as specific ground ring states
 \LZ \witteng\
(with ghost number one and zero, respectively). In the
conventional notation we can write
them as $\ti{a}{\sl O}_{s,s}$
and  ${\sl O}_{s,s}$, where $\ti{a}\equiv c\gamma$.

\medskip
\medskip
\medskip

To get the complete spectrum \LZ , we also need to combine the LWS
in $D^+_{j}$ in addition to the HWS in $D^-_{j}$ when we first
start with the Ramond state description \srtsd\ in the untwisted
theory as in \MKV.
Starting with $\Phi_{-j,j}$ in the Ramond state with the
opposite spin $e^{-\f{i}{2}H}$, after the spectral flow we can
find \eqn\duart{ T_{j}=c\de c\
e^{\s{\f{2}{n}}[i(j+\f{n}{2}-1)X+(j+\f{n}{2})\phi]}
\delta(\beta),} where $\delta(\beta)$ is defined by
$e^{\ti{\phi}}$ in terms of the bosonized field of $\beta\gamma$
system (see the appendix A). The operators \duart\ have the ghost
number two and are one to one correspondence to the ghost number
one counterpart $V_{j-1}$ in the two dimensional string theory. We
can also find by acting $(J^+_0)^{-2j+1}$ for $j\leq 0$ on \duart\
\eqn\tttf{\ti{T}_{j}=c\de c\ (\beta)^{2j-2}
e^{\s{\f{2}{n}}[-i(j-\f{n}{2})X+(j+\f{n}{2})\phi]},} which are the
ghost number two counterpart of $\ti{V}_{j-1}$. Thus we can obtain
the tachyon states\foot{ To be correct, we cannot find
$\ti{a}Y^{-}_{s,-s}$ because of the bound $j\leq 0$. It comes from
the previous HWS tachyon state $Y^-_{s,s}=V_{ns-1/2}$ by acting
$\ti{a}(K^-)^{2s}$.}
 at ghost number two
(i.e. $\ti{a}Y^{\pm}_{s,\pm s}$) in the $c<1$ string as $T_{j}$ and
$\ti{T}_{j}$.

We can show the relation $T_{j}\sim\ti{T}_{-j+1}$ from \eqivr\ as
before. Indeed there are non-vanishing two point functions $\la
V_{j}T_{-j}\lb$, $\la \ti{V}_{-j-1}T_{-j}\lb$, $\la
V_j\ti{T}_{j+1}\lb$ and $\la \ti{V}_{-j-1}\ti{T}_{j+1}\lb$ as is
clear from the reflections due to Liouville wall in the $c<1$
string. They can also be computed by using the known expression of
the two point functions in $SL(2,R)$ WZW model as we will discuss in
the section 3.2.

\medskip
\medskip
\medskip

To find the discrete states \LZ\ in the $c<1$ string, we can again
apply the method taken in the $c=1$ string case \MKV. Define the
operator called $K^-$ by boosting our model into the $c=1$ string
background \eqn\koo{K^{-}=\beta e^{-i\s{2}X'},} where $X'$ is the
Euclidean time in the $c=1$ string. It is defined by the boost
\eqn\boost{\eqalign{
iX'&=\f{i}{2}(\s{n}+1/\s{n})X+\f{1}{2}(\s{n}-1/\s{n})\phi,\cr
\phi'&=\f{i}{2}(\s{n}-1/\s{n})X+\f{1}{2}(\s{n}+1/\s{n})\phi.}} By
acting $K^-$ on the previous tachyon states and the ground ring
states, we can find all physical states\foot{ The definition of
$Y^{\pm}_{s,n}$ and ${\sl O}_{s,n}$ are given by in terms of $c=1$
fields (see \MKV ) $Y^{\pm}_{s,n}=c(K^-)^{s-n}\exp(i\s{2}X'
+\s{2}(1\mp s)\phi ')$ and ${\sl O}_{s,n}
=(K^-)^{s-n}\gamma^{2s}\exp(i\s{2}X'-\s{2}\phi')
=\beta^{-s-n}x^{s+n}y^{s-n}$, where $x$ and $y$ is the standard
ground generators.}
 in the $c<1$ string including
the discrete states and ground ring states
$Y^{\pm}_{s,n},\ti{a}Y^{\pm}_{s,n}, {\sl O}_{s,n}$ and
$\ti{a}{\sl O}_{s,n}$. Notice that the action of $K^-$ on $V_{j}$
is only well-defined when $2j+1 \in nZ$ due to the requirement of
 the locality of their OPE.

In this way we have checked that the physical spectrum of the
twisted $SL(2,R)_{2+n}/U(1)$ model (or its $Z_n$ orbifold)
coincides with that of the non-minimal $c<1$ string compactified
at the radius $R=\s{\f{2}{n}}$ (or $R=\s{2n}$).

\newsec{Three Point Functions}

Now we would like to move on to the analysis of interactions in
the twisted coset theory. We will compute the two and three point
functions and check that they are matching with the scattering
amplitudes in the $c<1$ string. What we have to show for the
consistency is that the correlation functions agree with each
other up to certain non-local factors (i.e. normalization) for
each vertex operators. Furthermore, as we will see later, these
momentum dependent factors are essentially the same as the
leg-factors (see e.g. \matreview) in the $c<1$ string computed in
\DK \TLD. As a specific limit,
 our results in this section
 also provide a new evidence for the analogous equivalence
in $c=1$ string \MKV.

\subsec{Two and Three Point Functions in $SL(2,R)$ WZW Model}

First we review the known results of two and three point functions
in $SL(2,R)_{k=n+2}$ WZW model \Ts \GK\ since they are the
essential parts of the correlation functions in the twisted coset
theory. We will follow the convention in \GK \GN. Notice that in
order to shift the convention in the previous section to the one
in this section, one has\foot{ This is because in \GK \GN, the
coupling constant is defined by $g_s=e^{-\f{Q}{2}\phi}$. On the
other hand, the field $X$ has the same sign as in the previous
section.} to change the sign: $\phi \to -\phi$. In this subsection
we write the primaries in bosonic $SL(2,R)/U(1)$ by
$V_{j,m,\bar{m}}$ (see \primarysl), where $m$ and $\bar{m}$ are
the eigenvalues of $J^3_0$ in the left and right-moving sector.
Note that the correlators of $V_{j,m,\bar{m}}$ are essentially the
same as those in the bosinic $SL(2,R)$ WZW model.

The non-trivial two point function of primaries are given by the
following formula\foot{Here we omitted the delta functions
$\delta(j-j')\delta^2(m+m')$ when we consider the two point
function for $V_{j,m,\bar{m}}$ and $V_{j',m',\bar{m}'}$.}
\eqn\twopp{\eqalign{&\la V_{j,m,\bar{m}}V_{j,-m,-\bar{m}}\lb\ \
(\equiv R(j,m,\bar{m})) \cr &=n(\nu)^{2j+1}\f{\Gamma(1-(2j+1)/n)
\Gamma(-2j-1)}{\Gamma(\f{2j+1}{n})\Gamma(2j+2)}\cdot
\f{\Gamma(j-m+1)\Gamma(1+j+\bar{m})}{\Gamma(-j-m)\Gamma(\bar{m}-j)},}}
where $\nu$ is defined by $\nu=\f{\Gamma(1+1/n)}{\pi
\Gamma(1-1/n)}$. Remember that in addition we have the trivial
ones $\la V_{j,m,\bar{m}}V_{-j-1,-m,-\bar{m}}\lb=1$. Notice also
that $R(j,m,m)$ is equal to the reflection coefficient in \eqivr.

The three point functions are given\foot{Again we suppressed the
delta functions $\delta^2(m_1+m_2+m_3)$.} by
\eqn\theesl{\eqalign{& \la
V_{j_1,m_1,\bar{m}_1}V_{j_2,m_2,\bar{m}_2} V_{j_3,m_3,\bar{m}_3}
\lb \cr & =\f{n}{(2\pi)^3} (\nu)^{j_1+j_2+j_3+1}
F(j_1,m_1,\bar{m}_1;j_2,m_2,\bar{m}_2;j_3,m_3,\bar{m}_3) \cr &
\times \f{G(-j_1-j_2-j_3-2)G(j_3-j_1-j_2-1)
G(j_2-j_1-j_3-1)G(j_1-j_2-j_3-1)}{G(-1)G(-2j_1-1)G(-2j_2-1)
G(-2j_3-1)},}} where the function $G(j)$ defined in \Ts \GK\
satisfies \eqn\formg{\eqalign{&G(j)=G(-j-n-1),\cr
&G(j-1)=\f{\Gamma(1+\f{j}{k})}{\Gamma(-\f{j}{k})}G(j), \cr
&G(j-n)=n^{-2j-1}\f{\Gamma(1+j)}{\Gamma(-j)}G(j). }} Another
function $F(j_i,m_i)$ is defined by the following multiple
integral \GK\ \eqn\funcff{\eqalign{&
F(j_1,m_1,\bar{m}_1;j_2,m_2,\bar{m}_2;j_3,m_3,\bar{m}_3)\cr &=\int
dx_1^2dx_2^2\ x_1^{j_1+m_1}\bar{x}_1^{j_1+\bar{m}_1}
|1-x_1|^{-2(j_1+j_2-j_3+1)}\cr &\ \ \times
x_2^{j_2+m_2}\bar{x}_2^{j_2+\bar{m}_2}
|1-x_2|^{-2(j_2+j_3-j_1+1)}|x_1-x_2|^{-2(j_1+j_2-j_3+1)}.}} In
general, it is difficult to represent $F(j_i,m_i)$ in terms of
simple functions. However, as shown in \HS, in the particular case
$j_1+m_1=j_1+\bar{m}_1=0$, we have a following useful expression
\eqn\formulaf{\eqalign{&
F(j_1,m_1,\bar{m}_1;j_2,m_2,\bar{m}_2;j_3,m_3,\bar{m}_3)\cr
&=(-1)^{m_3-\bar{m}_3}\pi^2
\f{\gamma(-j_1-j_2-j_3-1)\gamma(2j_1+1)}
{\gamma(1+j_1+j_2-j_3)\gamma(1+j_1+j_3-j_2)}\cdot
\f{\Gamma(1+j_2+m_2)\Gamma(1+j_3+m_3)}
{\Gamma(-j_2-\bar{m}_2)\Gamma(-j_3-\bar{m}_3)},}} where
$\gamma(x)\equiv \Gamma(x)/\Gamma(1-x)$. For example, if we set
$j_1=m_1=\bar{m}_1=0$ in \theesl\ and apply the formula \formulaf
, then we can check that the three point functions are
reduced\foot{Here we need to replace the divergence
$\f{\Gamma(0)}{2\pi}$ with $\delta(j-j)$.} to the two point
function \twopp.

\subsec{Two and Three Point Functions in Twisted Coset Theory}

We would like to compute the two and three point functions of
tachyon vertex operators $V_j, \ti{V}_j, T_j$ and $\ti{T}_j$. The
essential parts of the correlation functions are obviously those
of the bosonic $SL(2,R)$ WZW model\foot{In this paper we do not
consider the winding number violating amplitudes discussed in e.g.
\Gr. The author would like to thank Gaston Giribet and Yu Nakayama
very much for pointing out this point. In this case, the fermion
number (or $H$ momentum) is conserved as we can check in all
examples discussed in this subsection.} given by \twopp\ and
\theesl. In the $D^-_j$ representation, the tachyon vertex
operators in the twisted coset model can be expressed in the
Ramond state description as follows\foot{In the $D^+_{-j-1}$
representation, we can find $V_j=\Phi_{-j-1,j},\ \
\ti{V}_j=\Phi_{-j-1,-j-1},\ \ T_{j+1}=\Phi_{-j-1,j+1}$ and
$\ti{T}_{j+1}=\Phi_{-j-1,-j}$.} \eqn\summg{V_j=\Phi_{j,j},\ \
\ti{V}_j=\Phi_{j,-j-1},\ \ T_{j+1}=\Phi_{j,j+1},\ \
\ti{T}_{j+1}=\Phi_{j,-j}.} To make a physical vertex in the closed
string we need to combine both the left and right-moving sector.
Since the momentum modes in the $c<1$ string side correspond to
the winding modes in the coset model, we can impose the
restriction $m=\bar{m}$. Thus the correlators of $\Phi_{j,m}$ are
equal to those of $V_{j,m,m}$. The operators $V_j, \ti{V}_j,
T_{j+1}$ and $\ti{T}_{j+1}$ are tachyon vertex operators in the
$c<1$ string with the $\phi$ momentum
$p_{\phi}=i\s{\f{2}{n}}(j+\f{1}{2})$ as can be seen\foot{Notice
that here we remembered the relation of the convention in this
section and the one in section 2, i.e. $\phi\to -\phi$.}
 from
\opb, \vtjtak, \duart\ and \tttf. The $X$ momentum of $V_j$ and
$T_{j+1}$ is $p_{X}=\s{\f{2}{n}}(j+\f{1}{2})$, while the momentum
of $\ti{V}_j$ and $\ti{T}_{j+1}$ is
$p_{X}=-\s{\f{2}{n}}(j+\f{1}{2})$.

Let us first compute the three point functions in the twisted
theory. To do this we can replace two of the three operators with
the Ramond ones and the other one with the same one as in the
twisted theory (i.e. NS operator) following the general principle
\WiTS. Consider the three point functions
$\ti{C}(j_1,j_2,j_3)\equiv\la \ti{V}_{j_1} \ti{V}_{j_2}
\ti{V}_{j_3} \lb$ in the twisted theory. They can be computed from
the three point functions in the untwisted theory
\eqn\cortopp{\ti{C}(j_1,j_2,j_3)= \la
\Phi_{j_1-\f{n}{2},-j_1+\f{n}{2}}\ \Phi_{j_2,-j_2-1}\
\Phi_{j_3,-j_3-1} \lb.} Indeed the $J^3_0$ charge conservation
$j_1+j_2+j_3=-2+\f{n}{2}$ agrees with the momentum conservation in
the $X$ direction \eqn\momcpq{p^1_X+p^2_X+p^3_X-i\f{Q(X)}{2}=0.}

Then we can find the explicit expressions of the three point
functions from the formula \theesl. Also the function $F(j_i,m_i)$
is simplified owing to the formula \formulaf. In the end, we find
\eqn\vtithree{ \ti{C}(j_1,j_2,j_3)=\f{\Gamma(0)^2}{2\pi\nu}
\cdot\gamma\left(1-\f{2j_1+1}{n}\right)\cdot
\gamma\left(1-\f{2j_2+1}{n}\right)
\cdot\gamma\left(1-\f{2j_3+1}{n}\right).}
This result remarkably agrees with the three point functions in
the $c<1$ string up to an overall (divergent) constant
factor\foot{ This factor is given by
$\f{\Gamma(0)^2}{2\pi}(\nu)^{(1-n)/n}$. Of course, it can be again
included in the non-local factors for the three vertex operators.
}. To see this, remember that the three point functions in the
$c<1$ string is just a constant except the non-local factor (or
the leg-factor) for each (incoming) particles $(i=1,2,3)$ \DK \TLD
 \eqn\legth{\mu^{-\s{\f{n}{2}}p^i_X}\cdot
\gamma\left(1+\s{\f{2}{n}}p^i_{X}\right)=
\nu^{j_i+1/2}\cdot\gamma\left(1-\f{2j_i+1}{n}\right).} Here we
identified the constant $\nu$ with the cosmological constant $\mu$
in the Liouville theory. This is because $\nu$ is proportional to
the coefficient of the screening operator \scr \GK \GN. The
non-local factor $\gamma(1+\s{\f{2}{n}}p^i_{X})$ is exactly the
same as the leg-factor\foot{ The leg-factor \legth\ in the $c<1$
string was first computed in \TLD\ by comparing string theory
S-matrix \DK\ with the matrix model dual \fermiss. At $b=1$ (or
$n=1$), it is obviously reduced to the familiar leg-factor
\matreview\ in the $c=1$ string.} in the $c<1$ string.

In the similar way we can analyze other three point functions. For
example, consider the three point functions $C(j_1,j_2,j_3)=\la
V_{j_1} V_{j_2} V_{j_3} \lb$ in the twisted theory. This is
essentially reduced to \eqn\cortopp{C(j_1,j_2,j_3)
=\la\Phi_{-j_1-\f{n+2}{2},j_1+\f{n+2}{2}}\ \Phi_{j_2,j_2}\
\Phi_{j_3,j_3} \lb,} where the first one corresponds to the \opb,
and the second and third one to the Ramond operator \srtsd. Indeed
the $J^3_0$ charge conservation $j_1+j_2+j_3+\f{n+2}{2}=0$
agrees with the momentum conservation \momcpq\ in the $c<1$
string. We can again compute the correlators by applying
 \theesl\ and \formulaf. It is given by
\eqn\vvvth{C(j_1,j_2,j_3)=\f{1}{2\pi n}
\cdot\nu^{-2j_1-n-1}\cdot\gamma\left(1+\f{2j_1+1}{n}\right).}
Since we can obviously express this by the three non-local factors
times a constant, it is consistent with the three particle
scattering in the $c<1$ string. However, one may worry about the
asymmetry in $\vvvth$ with respect to the permutation of each
particle. This is resolved if we replace the operators
$\Phi_{-j_1-\f{n+2}{2},j_1+\f{n+2}{2}}$ with the dual ones
$\Phi_{j_1+\f{n}{2},j_1+\f{n+2}{2}}$ using the equivalence \eqivr.
 Then the
three point functions simply become
\eqn\vvvthr{C'(j_1,j_2,j_3)\equiv
\la\Phi_{j_1+\f{n}{2},j_1+\f{n+2}{2}}\ \Phi_{j_2,j_2}\
\Phi_{j_3,j_3} \lb= \f{n\Gamma(0)}{2\pi},} where we applied the
relation \eqivr\ to \vvvth. Indeed we can see this is symmetric as
in the $\ti{V}_j$ case \vtithree. One interesting feature is that
this time we do not have any momentum dependent factor like
\legth. This is not surprising because the vertex $V_{j}$ has the
opposite chirality to $\ti{V}_{j}$ and they can have different
normalizations. This suggests that we should adjust the
normalization of $V_j$ and others so that the non-local
transformation for the $\ti{V}_{j}$ does not depend on the
momentum. In this case we have the complete agreement with the
$c<1$ string theory including the leg factors.

 Two point functions in the topologically twisted theory
are the same as those of the corresponding Ramond operators in the
untwisted theory. We can find the nontrivial two point functions
(reflection amplitudes) from \twopp\ \eqn\corgp{\la
\ti{V}_{j}T_{j+1}\lb =R(j,-j-1,-j-1)\   ,\ \ \  \la
V_j\ti{T}_{j+1}\lb =R(j,j,j)\ , } as well as the trivial ones
\eqn\corgpt{\la V_{j}T_{-j}\lb =\la \ti{V}_{j}\ti{T}_{-j}\lb =1.}
Here $R(j,m,m)$ are explicitly given by
 \eqn\twopor{\eqalign{
&R(j,-j-1,-j-1)=n\nu^{2j+1}\Gamma(0)
\cdot\gamma\left(1-\f{2j+1}{n}\right),\cr
&R(j,j,j)=\f{\nu^{2j+1}}{n\Gamma(0)}\cdot
\gamma\left(-\f{2j+1}{n}\right).}} These two point functions again
are consistent with those in the $c<1$ string up to non-local
factors (or leg-factors). The dependence of $\nu(=\mu)$ agrees
with the scaling of $\mu$ in the Liouville theory \DDK. The first
one in \twopor\ looks consistent with the \legth; i.e. the factor
$\nu^{j+1/2}\Gamma(0)\cdot\gamma\left(1-\f{2j+1}{n}\right)$ can be
regarded as the non-local factor for the vertex $\ti{V}_{j}$
(compare this with \vtithree) . The one\foot{Even though the
non-local factors we found for $\ti{V}_j$ and $T_j$ are
asymmetric, we do not think this is problematic just because
$\ti{V}_j$ and $T_j$ have different ghost numbers and their
origins are also completely different as is clear from section
2.4.} for $T_{j+1}$ is $n\nu^{j+1/2}$. We can also have the
agreement by assigning the factor
$\f{\nu^{j+1/2}}{n\Gamma(0)}\gamma\left(-\f{2j+1}{n}\right)$ and
$\nu^{j+1/2}$ to $\ti{T}_{j+1}$ and $V_j$, respectively.

\newsec{Topological Landau-Ginzburg model}

Until now, we have checked the physical spectrum and the three
point functions in the twisted $SL(2,R)_{n+2}/U(1)$ model agree
with those in the non-minimal $c<1$ string. To compute more
general interactions , i.e. $n(>3)$ point functions in the coset
model side, it is one of the easiest way to utilize its
topological LG model description. The twisted $SL(2,R)_{2+n}/U(1)$
model is expected to be equivalent to
 the topological LG model
 with the potential
\eqn\LGP{W=-\f{\mu}{n}X^{-n},} as conjectured in \OV. It can also
be regarded as an extension of the well-known relation between the
$N=2$ $SU(2)_{k}/U(1)$ coset and the $N=2$ LG model with potential
$W=X^{k+2}$ to the negative values of $k$. It will also be useful
to notice that the twisted $SU(2)_{k}/U(1)$ model or equally the
twisted $N=2$ minimal model \EY\ is known to be equivalent to the
minimal $(1,n)$ string \KL \DVV. The relation between the twisted
coset model and the LG model \LGP\ can also be understood from the
mirror symmetry or a supersymmetric version of FZZ duality \HoKa \Nak.
In this viewpoint the LG model is the B-model
mirror description for the A-model topological string on
$SL(2,R)_{n+2}/U(1)$ defined in section 2.2.

Below we will calculate various tree level interactions in the
topological LG model \LGP\ and check that it agrees with the
scattering amplitudes in the $c<1$ string. For $n=1$ case the
computations were performed in \GM \HO\ and they agree with those
in the $c=1$ string. Since our analysis here is a natural
extension of these results, the discussions in this section will
be very brief. In the end, our results in this section again
provide a further evidence that supports the equivalence of the
twisted coset model and the non-minimal $c<1$ string.

\subsec{Scattering Amplitudes}

We can find the following correspondence between a closed string
tachyon operator $T_k$ and a operator in the twisted LG theory
\eqn\tacco{T_{k}=X^{k-n},} where $k$ is an arbitrary integer and
is proportional to the momentum. It has the $U(1)$ ghost charge
$q_{k}=1-k/n$. The momentum conservation is given by
\eqn\momc{\sum_{i=1}^{N}k_i=2(n-1).} This comes from the $U(1)$
ghost charge conservation $\sum_{i}(q_{k_i}-1)=(g-1)(3-\hat{c})$
in the topological gravity description \KL \MKV. Then we get the
three point function (setting $\mu=1$) by applying the standard
residue formula in the topological LG model \DVV \eqn\threepp{ \la
T_{k_1}T_{k_2}T_{k_3}\lb =\int dX\ \f{T_{k_1}T_{k_2}T_{k_3}}{W'}=
\delta_{k_1+k_2+k_3-2n+2,0}.}

In order to compute the four point function, we need to take into
account of the contact terms \Lo \EKYY \eqn\contg{
C_{W}(T_k,T_{k'})=\f{d}{dX}\left(T_{k}T_{k'}/W'\right)_-
=(k+k'+1-n)\theta(-k-k'+n-1)T_{k+k'},} where the symbol $_-$ means
that we take only the negative power part \GM \HO. Thus the four
point function is given by the following expression including the
contributions from contact terms \eqn\fourp{\eqalign{& \la
T_{k_1}T_{k_2}T_{k_3}T_{k_4}\lb \cr &=\f{\de}{\de t_4}\la
T_{k_1}T_{k_2}T_{k_3} \lb_{W+t_4T_{k_4}} +\la
C_{W}(T_{k_4},T_{k_1})T_{k_2}T_{k_3} \lb +\la
T_{k_1}C_{W}(T_{k_4},T_{k_2})T_{k_3} \lb \cr & \ \ +\la
T_{k_1}T_{k_2}C_{W}(T_{k_4},T_{k_3}) \lb \cr
&=\delta_{k_1+k_2+k_3+k_4-2n+2,0} \cr &\ \ \  \times  \f{1}{2}
\left[(n+1)-|k_1+k_4+1-n| -|k_2+k_4+1-n|-|k_3+k_4+1-n|\right].}}
It is easy to check that this expression \fourp\ is symmetric
under any exchange of the four particles. When we restrict \fourp\
to the kinematical region of $1\to 3$ scattering \DK\ i.e.
$k_{2,3,4}>n-1$ and $k_1<n-1$, we find the following simplified
result \eqn\fourpf{\la T_{k_1}T_{k_2}T_{k_3}T_{k_4}\lb
=(k_1+1)\delta_{k_1+k_2+k_3+k_4-2n+2,0}.}

Furthermore, in the same way, we can compute the five particle
scattering amplitudes by perturbing potential infinitesimally. For
the kinematical region of $1\to 4$ scattering, we obtain\foot{To
see this we note that the five point function can be reduced to
four point functions via the derivative of the perturbed ones as
in \fourp. This is correct when the perturbation is a primary
field and it requires $k>0$. Indeed, the tachyon fields
$T_{k_{2,3,4}}$ satisfy this constraint. Also one more useful fact
is that the contact term only appears between the operators
$T_{k_{2,3,4}}$.} \eqn\fivep{\eqalign{& \la
T_{k_1}T_{k_2}T_{k_3}T_{k_4}T_{k_5}\lb \cr
&=\delta_{k_1+k_2+k_3+k_4+k_5-2n+2,0}\cdot (k_1+1)(k_1+n+1).}}

In general, we get \eqn\genrrr{\la T_{k_1}T_{k_2}\ddd T_{k_N}\lb
=\left(n\f{d}{d\mu}\right)^{N-3}\mu^{-(k_1+1)/n},} making the
$\mu$ dependence explicit. Though the momentum dependent factor
$\mu^{-(k_1+n-1)/n}$ is missing in the above computation, we can
see that it comes from the explicit identification of negative $k$
states $T_k$ with the gravitational descendants\foot{Here we may
identify them with the descendant of the dual cosmological
operator so that the power of $\ti{\mu}=\mu^{1/n}$ is integer.} as
in the $n=1$ case \GM \HO.

Now let us compare this with the string theory results. The
relation between $p_X$ in the $c<1$ string (we assume $\al=2$ as
in section 3.2) \DK \TLD\ and the integer $k$ can be found from
the momentum conservation \momc\ as follows
\eqn\reek{p_X=\f{k}{\s{2n}}.} This relation can also be understood
from the comparison of cosmological constants $|W|^2\sim \mu
e^{\s{2n}\phi}$, which means\foot{Even though we used the same
symbol $X$ for both the Euclidean time (in section 2) and the LG
field (in section 3), which are completely different from each
other.} $|X|\sim e^{-\phi/\s{2n}}$. Then we can identify the
compactification radius with $R=\s{2n}$. Assuming $b=\s{n}$, we
can check that the previous amplitudes in the LG theory agree with
the string theory \DK\ or matrix model results \TLD\ up to the
following energy dependent factor (or leg-factor)
 for each particle\foot{One way to find this factor in the LG theory 
is to fix the 
overall normalization of the two point functions by rewriting 
them in terms of disk amplitudes. They are computed in \HIV\ for the
$SU(2)/U(1)$ case. Its extension to our $SL(2,R)/U(1)$ can be done via 
the simple continutation of the level $n\to -n$. The author would like to 
thank Cumrun Vafa very much for explaining this point}.
\eqn\legfh{\f{\Gamma(1-\s{\f{2}{n}}p_X)}{n\cdot\Gamma(\s{\f{2}{n}}p_X)}.}
In this way we have shown that the topological LG model \LGP\
describes the $c<1$ non-minimal string ($b=\s{n}$) at the radius
$R=\s{2n}$.

In this computation, as we have seen, the momentum dependent
factor or the leg factor \legfh\ in the $c<1$ string is what we
should put by hand. This was also true for the computation of the
three point functions for $V_j$ \cortopp\ obtained from the direct
computations in the coset model. These are not problematic since
the factor can be removed by a field redefinition for the
particle. Nevertheless, it would be intriguing to notice again
that the the three point functions for $\ti{V}_j$ \vtithree\
include the same non-local factor as the one in the $c<1$ string.
This suggests that we can determine the field redefinition by
adjusting the other operators to the canonical one $\ti{V}_j$.

\newsec{Summary and Discussions}

\subsec{Equivalence for Positive Integer $n$}

We have shown the following equivalence\foot{ Here again we assume
$\al=2$.} between the non-minimal $c<1$ string and the twisted
$N=2$ $SL(2,R)/U(1)$ model or the equivalent topological LG model
for integer $n$ \eqn\mailcc{\eqalign{& {\rm Nonminimal}\
c=1-6(n-1)^2/n\ \ \ {\rm String\ at\ radius}\ R=\s{2n} \cr &
\simeq {\rm Twisted\ A-model\ on}\
\left(\f{SL(2,R)_{2+n}}{U(1)}\right)/{Z_n} \ \ ({\rm radius}\
R_A=\s{\f{2}{n}}) \cr
 & \simeq {\rm Topological\ LG\ model}\  W=-\f{\mu}{X^{n}},}}
where the $Z_n$ projection is the translation by
$\f{2\pi}{\s{n/2}}$ in the circle direction of the cigar (see
e.g.\HoKa). Notice that the topological (B-model) twisted LG model
in \mailcc\ is expected to be equivalent to that of the $N=2$
Liouville theory via the supersymmetric version of the FZZ duality
or equally the mirror symmetry \HoKa. The main point is the
relation between the $c<1$ string and either of the two
topological models.
 Notice that in the above
model \mailcc\ the compactification radius $R=\s{2n}$ is
consistent\foot{Refer also to \IM\ for a similar compactification
in the presence of an imaginary linear dilaton gradient. In this
case the time-like coordinate is compactified, while in our case
the space-like coordinate $X$ is so.} with the string coupling
constant $g_s=e^{i(\s{n/2}-1/\s{2n})X}$.

We have checked this claim by computing the physical spectrum and
the scattering amplitudes (or equally correlation functions) in
the twisted theories and comparing them with those in the $c<1$
string. In particular we directly calculated the three point
functions in the twisted $SL(2,R)/U(1)$ model in terms of the
untwisted theory. This reveals the structure of the momentum
dependent factor (or leg factor) for each tachyon vertex. We found
that this also essentially agrees with that of the $c<1$ string.

\subsec{Equivalence in More General Cases}

Up to now, in the most of the discussions, we have assumed that
$n$ is a positive integer. One important subtle point in the non
integer case is that the coupling constant dependence
$g_s=e^{i(\s{n/2}-1/\s{2n})X}$ does not seem to respect the
periodicity of the compactification radius $R=\s{2n}$. However,
many results seem to make sense even if $n>0$ is not an integer.
For example, the computation of the physical spectrum in section 2
and the scattering amplitudes in section 3 and 4 can be done for
general $n$ and we get the same results. Thus it is natural to
think that we can continuously change the value of level $n$ of
the $SL(2,R)/U(1)$ coset. These observations suggest that we may
be able to define the $c<1$ string at the specific radius by the
topological twist of
 $SL(2,R)/U(1)$ model in the following way
\eqn\maildd{\eqalign{& {\rm Nonminimal}\  c=1-6(n-1)^2/n\ \ \ {\rm
String\ at\ radius}\ R=\s{\f{2}{n}} \cr & \simeq {\rm Twisted\
A-model\ on}\ \f{SL(2,R)_{2+n}}{U(1)} \ \ ({\rm radius}\
R_A=\s{{2n}}).}} We can also take the $Z_{m}$ ($m$ is a positive
integer) identification in the circle direction; the special case
$n=n'$ will be reduced to \mailcc. Also notice that when $n$ is an
integer the model defined in \maildd\ is also equivalent to the
$Z_n$ quotient of the LG model $(W=-\f{\mu}{X^{n}})/{Z_n}$. In
this case the $Z_n$ projection restricts the tachyon operators
$T_{k}=X^{k-n}$  to the particular momenta $k\in nZ$.

When $n$ is rational i.e. $n=\f{p}{q}$ in terms of the coprime
integers $p,q$, the $c<1$ string has the same central charge as
the minimal $(p,q)$ model. Thus we may call the model a
non-minimal $(p,q)$ model. For example, if we consider the
(critical) 2d black hole in type 0 string (i.e. $n=1/2$) and
perform the topological twist, then the result is equivalent to
the non-minimal $(1,2)$ model.

The exchange $n\lr 1/n$ in the twisted $SL(2,R)/U(1)$ theory does
not change the matter central charge in its equivalent $c<1$
string theory while the radius $R$ is replaced with $nR$. In
particular, we can find that the twisted $SL(2,R)_{2+1/n}/U(1)$
model is equivalent to the twisted $(SL(2,R)_{2+n}/U(1))/Z_n$
model in \mailcc\ when $n$ is an integer.
 This relation comes from the
$b\to 1/b$ duality of the Liouville sector in the $c<1$ string
\Dual. This should be related to the supersymmetric FZZ-duality
\HoKa\ that the $N=2$ Liouville term  (when its absolute valued
square is taken, its $\phi$ dependence is $\sim e^{\s{2n}\phi}$)
and the $SL(2,R)$ screening operator ($\sim e^{\s{\f2n}\phi}$) are
dual to each other and this corresponds to the $b\to 1/b$ duality
in the $c<1$ string.

It will also be intriguing to consider the case $n<0$. The similar
problem in the (untwisted) bosonic $SL(2,R)/U(1)$ CFT was
discussed in \HiTa\ from the viewpoint of time-dependent
background in string theory. In the non-critical string
description of our model, the matter sector in the world-sheet
theory is described by the time-like Liouville theory \ST \SCHM\
because we now have the screening potential in the time-direction
 after the Wick rotation \TLD. Thus the model becomes very close
to the minimal model. Indeed when $n$ is a negative integer, the
LG potential becomes $W= X^{|n|}$. Thus it is the same as the
twisted $N=2$ minimal model (or the coset $SU(2)_n/U(1)$), which
is known to be equivalent to the minimal $(1,|n|)$ string or the
topological gravity coupled to the $|n|-$th minimal matter \KL
\DVV \OV \ADKMV.

To find analogous twisted SCFT models, such as some similar coset
models, which are equivalent to the $\hat{c}=1$ or $\hat{c}<1$
type 0 string \TT \six \KMSS\ will be another interesting problem
(refer to e.g.\OS\ for recent relevant discussions). An important
new ingredient in this case will be the existence of the RR-flux
backgrounds \six \Kap \Dan \GTT. We leave this as a future
problem.

We would also like to point out a quite recent paper \RT, which
appeared on the web after we finished the main computations in the
present paper. In this paper, an explicit relation has been found
between the correlation functions in the bosonic $SL(2,R)$ WZW
model (for arbitrary values of the level) and those in the
Liouville theory. This seems to be closely related to our relation
between the $N=2$ twisted coset model and the $c<1$ string since
the relevant Liouville theory looks identical. It may help us to
prove our claim for more general correlation functions, though we
will leave the detailed study of this issue for a future
publication.

\subsec{Matrix Model Dual and Landau-Ginzburg Potential}

The matrix model dual of the $c<1$ string is described by the
fermi surface (setting $b=\s{n}$ in \fermiss) \TLD
\eqn\fermidefor{(W_{(1,0)})^n\cdot W_{(0,1)}=\mu,} where
$W_{(1,0)}=(-p-x)e^{-t}/2$ and $W_{(0,1)}=(p-x)e^{-t}/2$ are the
elementary conserved charges.

When $n$ is an positive integer, the circle compactification in
the topological models corresponds to the periodicity $t\sim
t+(1+n)\pi i$ in the time-direction\foot{ The imaginary factor $i$
means that the compactification should be taken in the Euclidean
time direction as we did in the $SL(2,R)/U(1)$ model.}. As can be
seen from \boost\ and the asymptotic behavior $x\sim e^{-\phi}$
\matreview, the compactification does not affect $W_{(1,0)}$ and
$W_{(0,1)}$. Thus the $c<1$ string , for an integer $n$, is
expected to be described by the time-dependent background
\fermidefor\ of $c=1$ matrix model with the compactified Euclidean
time coordinate (radius\foot{Notice that in the matrix model side
we assume $\al=1$ following the standard notation.}
$R_M=\f{1+n}{2}$).

When $n$ is a rational number
 $n=p/q$, then the matrix model dual seems to be a more
 complicated one. If we follow the same logic as before, then
it is described by the fermi surface \fermidefor\ with the $Z_q$
identification $t\sim t+\f{p+q}{p}\pi i$ and
$(W_{(1,0)},W_{(0,1)})\sim (W_{(1,0)},\ e^{2\pi iq/p}W_{(0,1)})$.
The explicit calculations of physical quantities in this quotient
matrix model will be an interesting future problem.

 On the other hand, at a formal level, we may apply the LG/CY
correspondence \WIL \OV\ to our previous results of the LG model.
Then we naively obtain the `$d=1+2/n$ dimensional surface'
($b=\s{n}$), \eqn\curveff{\mu X^{-n}+\sum_{i=1}^{1+1/n}Y_iZ_i=0,}
in the weighted projective space $WP_{(-2,n,n,\ddd)}$. It is easy
to see that the space \curveff\ has a vanishing first Chern-class.
When $n=1$, this space \curveff\ is the same as the conifold as is
well-known \OV. Even though generally $d$ is fractional, we can
obviously see that the LG model corresponding to \curveff\ is
indeed the same as \LGP. If we consider fractional numbers for $n$
as in section 5.2, $d$ can be an integer. If we set $Y_1=1$ by
gauge fixing and neglect `winding modes' $Y_i,Z_{i}\ \
(i=2,3,\ddd)$, then we can find $X^nZ_1=\mu$, which is the same as
\fermidefor. We can think this is the natural extension of the
similar result \OV\ for $n=1$ or $c=1$ string.

Finally, it will be very helpful to uncover any relation between
the critical 2d black hole in type 0 string and its twisted
version as the ten dimensional type II string on Calabi-Yau space
is related to the topological string on the same space. The
twisted coset model is equivalent to the non-minimal $(1,2)$ model
or equally $c=-2$ string as we have argued in section 5.2. The
matrix model dual of the 2d type 0 black holes is already proposed
\giveon \Xi \PaSu\ based on the bosonic string counterpart \KKK.
Thus we can compare it with the matrix model dual for the twisted
theory which we have found just before. This will reveal the
meaning of twisting in terms of the matrix model. On the other
hand, a seemingly related equivalence has recently been found in
\GRT\ that the $\hat{c}=0$ type 0 string \KMSS\ is equivalent to
the $Z_2$ orbifold of $c=-2$ (bosonic) string. This issue will
also be an important future direction.

\vskip 5in

\centerline{\bf Acknowledgments}

 \vskip .1in

 I would like to thank G. Giribet,
Y. Hikida, Y. Nakayama, Y. Sugawara, H. Takayanagi, S.
Yamaguchi and C. Vafa 
very much for stimulating discussions and enlightening
explanations. I am also grateful to A. Adams, J. L. Karczmarek, J.
McGreevy, T. Nakatsu, N. Ohta, T. Sakai, D. Shih, A. Strominger,
S. Terashima, A. Tsuchiya and M. Wijnholt for helpful
discussions and comments. I would like to thank the Center of
Mathematical Sciences in Zhejiang University for its hospitality.
This work was supported in part by DOE grant DE-FG02-91ER40654.

\vskip .3in

\appendix{A}{Spectral Flow in $SL(2,R)$ WZW model}
Here we give a brief review of the spectral flow in the $SL(2,R)$
WZW model \MO\foot{Notice that the convention of this paper is
different from \MO\ by the sign $j_{MO}=-j_{us}$. Also the one for
\GK\ is such that $j_{GK}=-j_{us}-1$.}. This is defined by the
following transformation with the winding number $w$ ($w\in {\bf
Z}$) which preserves the bosonic $SL(2,R)$ (level $k=n+2$) current
algebra \eqn\specrtf{J^{3'}_{n}= J^3_{n}+\f{n+2}{2}w\delta_{n,0},\
J^{\pm'}_{n}= J^{\pm}_{n\mp w}.} In the notation in section 2, it
is equivalent to the shift $|0\lb \to
e^{\s{\f{n+2}{2}}wX_3}|0\lb$. Then the Virasoro operators shift
\eqn\visf{L_n'=L_n-wJ^3_{n}-\f{n+2}{4}\delta_{n,0}.} It is also
easy to see explicitly (by plotting the allowed values of
$(L_0,J^3_0)$) the equivalence \eqn\eqi{ D^{\pm,w=\mp 1}_{j}\simeq
D^{\mp,w=0}_{-j-\f{n+2}{2}}\ \ \ .}

In the Wakimoto representation the spectral flow corresponds to
\eqn\wakisp{ \beta'_{n}= \beta_{n+w},\ \ \
\gamma'_{n}=\gamma_{n-w},\ \ \
\de\phi'=\de\phi-\f{n+2}{\s{2n}}\f{w}{z}.} Its action on a state
is \eqn\actsp{|0\lb_{\phi} \to e^{\s{\f{n}{2}}w\phi}|0\lb,\ \ \
|0\lb_{\beta\gamma}\to |w\lb_{\beta\gamma},} where
$|w\lb_{\beta\gamma}$ is defined by
\eqn\dedfd{\beta_{n-w}|w\lb_{\beta\gamma}=\gamma_{n+1+w}
|w\lb_{\beta\gamma}=0,\ \ \ (n=0,1,2,\ddd).} If we use the
bosonized representation \eqn\bosobg{\beta=e^{-\ti{\phi}}\de\xi,\
\ \ \gamma=e^{\ti{\phi}}\eta,} the shifted of vacuum is expressed
as \eqn\shiftggg{|w\lb_{\beta\gamma}=e^{w\ti{\phi}}|0\lb.} The
conformal dimension of the operator $e^{l\ti{\phi}}$ is $\Delta
=-l(l+1)/2$.

\listrefs

\end